\documentclass[aps,prd,floatfix,nofootinbib,twocolumn,10pt]{revtex4}
\usepackage{color,amsmath,amssymb,graphicx,latexsym,subfigure}
\usepackage{threeparttable,txfonts}
\usepackage{url}

\makeatother

\newcommand{\eV}{{\rm eV}}

\allowdisplaybreaks[2]

\begin{document}

\title{Searching for axion dark matter with the MeerKAT radio telescope}

\date{\today}

\author{Yun-Fan Zhou$^{1,2}$, Nick Houston$^3$, Gyula I. G. J\'{o}zsa$^{4,5,6}$, Hao Chen$^{7,1,8}$,
Yin-Zhe Ma$^{9,10,11,1}$\footnote{Corresponding author: Y.-Z. Ma, ma@ukzn.ac.za}, 
Qiang Yuan$^{12,2}$\footnote{Corresponding author: Q. Yuan, yuanq@pmo.ac.cn}, 
Tao An$^{13}$, Yogesh Chandola$^1$, Ran Ding$^{14}$, Fujun Du$^{1,2}$, Shao-Guang Guo$^{13}$, 
Xiaoyuan Huang$^{12,2}$, Mengtian Li$^{12,2}$, Chandreyee Sengupta$^1$}

\affiliation{
$^1$Key Laboratory of Radio Astronomy, Purple Mountain Observatory, Chinese Academy of Sciences, 
Nanjing 210023, China\\
$^2$School of Astronomy and Space Science, University of Science and Technology of China, Hefei, Anhui 230026, China \\
$^3$Institute of Theoretical Physics, Faculty of Science, Beijing University of Technology, Beijing 100124, China\\
$^4$Max-Planck-Institut f\"ur Radioastronomie, Radioobservatorium Effelsberg,
Max-Planck-Stra{\ss}e 28, 53902 Bad M\"unstereifel, Germany\\
$^5$Department of Physics and Electronics, Rhodes University, PO Box 94, Makhanda, 6140, South Africa\\
$^6$South African Radio Astronomy Observatory, Black River Park, 2 Fir Street, Observatory, Cape Town, 7925, South Africa\\
$^7$Department of Astronomy, University of Cape Town, Private Bag X3, 7701 Rondebosch, South Africa\\
$^8$Research Center for Intelligent Computing Platforms, Zhejiang Laboratory, Hangzhou 311100, China\\
$^9$School of Chemistry and Physics, University of KwaZulu-Natal, Westville Campus, Durban, 4000, South Africa\\
$^10$NAOC-UKZN Computational Astrophysics Centre (NUCAC), University of KwaZulu-Natal, Durban, 4000, South Africa\\
$^{11}$National Institute for Theoretical and Computational Sciences (NITheCS), South Africa \\
$^{12}$Key Laboratory of Dark Matter and Space Astronomy, Purple Mountain Observatory, Chinese Academy of Sciences, Nanjing 210023, China\\
$^{13}$Shanghai Astronomical Observatory, Chinese Academy of Sciences, Nandan Road 80, Shanghai 200030, China\\
$^{14}$School of Physics and Optoelectronics Engineering, Anhui University, Hefei 230601,China}

\begin{abstract}
\noindent

Axions provide a natural and well-motivated dark matter candidate, with the capability to convert directly to photons in the presence of an electromagnetic field. 
A particularly compelling observational target is the conversion of dark matter axions into photons in the magnetospheres of highly magnetised neutron stars, which is expected to produce a narrow spectral peak centred at the frequency of the axion mass. 
We point the MeerKAT radio telescope towards the isolated neutron star J0806.4$-$4123 for $10$-hours of observation and obtain the radio spectra in the frequency range $769$-$1051$\,MHz. 
By modelling the conversion process of infalling axion dark matter (DM), we then compare these spectra to theoretical expectations for a given choice of axion parameters.
Whilst finding no signal above $5\sigma$ in the data, we provide a unique constraint on the Primakoff coupling of axion DM, $g_{{\rm a}\gamma\gamma}\lesssim 9.3 \times 10^{-12}\,{\rm GeV}^{-1}$ at the $95\%$ confidence level, in the mass range $3.18$-$4.35\,\mu$eV. 
This result serves the strongest constraint in the axion mass range $4.20$-$4.35\,\mu$eV.

\end{abstract}

\maketitle

{\bf Introduction.}
As a minimal extension of the Standard Model, and in particular the Peccei-Quinn solution of the strong CP problem \cite{Peccei:1977ur, Weinberg:1977ma,Wilczek:1977pj,Vafa1984}, axions and axion-like particles occupy a rare point of convergence in theoretical physics, in that they are also a generic prediction of the exotic physics of string and M theory  \cite{Svrcek:2006yi, Arvanitaki:2009fg}.
Despite the profound differences between these contexts, their resultant properties are also largely universal, creating an easily characterisable theoretical target.

As typically light, long-lived pseudoscalar particles, axions are also natural candidates for the mysterious dark matter (DM) comprising much of the mass of our observable universe \cite{Dine:1982ah, Preskill:1982cy}, becoming a topic of intense ongoing investigation \cite{Marsh:2015xka,Irastorza:2018dyq}. 
In recent years a particularly compelling observational mechanism has emerged, Primakoff conversion of these DM axions into photons in the magnetospheres of highly magnetised neutron stars (NS)~\cite{Pshirkov:2007st}. 
Following Ref.~\cite{Pshirkov:2007st} this idea was revisited more thoroughly in Refs.~\cite{Hook:2018iia,Huang:2018lxq}, leading to a flurry of theoretical activities ~\cite{Safdi:2018oeu,Battye:2019aco,Leroy:2019ghm,Witte:2021arp,Battye:2021xvt,Millar:2021gzs}. 

Based on the observed DM density, the so-called QCD axions which solve the strong CP problem are favoured to have masses in the $10$-$100\,\mu{\rm eV}$ range ~\cite{Abbott:1982af}. 
This corresponds to a characteristic signature of NS conversion in the MHz-GHz frequency range, exactly covered by mid-frequency radio telescopes (e.g. Square Kilometre Array~\cite{Colafrancesco2015}, and FAST~\cite{Nan2011}). 
Radio antennae can therefore provide a unique and complementary method to laboratory axion DM haloscope experiments (e.g. ADMX~\cite{ADMX:2018gho}). Several observational studies recently have been performed using the Karl G. Jansky Very Large Array \cite{Darling:2020plz,Darling:2020uyo,Battye:2021yue}, and the Green Bank and Effelsberg radio telescopes \cite{Foster:2020pgt, Foster:2022fxn} to search for the signature of this conversion mechanism.


In the following we will also search for this phenomenon by observing the isolated NS J0806.4$-$4123 with the MeerKAT radio telescope~\cite{2016mks..confE...1J}. MeerKAT has a good sensitivity along with a wide UHF band coverage to search for this type of axion signature in the low mass window where there is a gap between ADMX \cite{ADMX:2021abc} and RBF \cite{DePanfilis} laboratory experiments.

{\bf Axion conversion in NS magnetospheres.}
Our starting point is the Lagrangian that represents coupling between the electromagnetic field and the axion field $a$, $\mathcal{L}_{{\rm a}\gamma}=-(1/4)g_{{\rm a}\gamma\gamma}aF_{\mu\nu}\tilde{F}^{\mu\nu}= g_{{\rm a}\gamma\gamma}a\vec E\cdot \vec B$, where $g_{{\rm a}\gamma\gamma}$ is the coupling constant and $\vec E/\vec B$ are the electric/magnetic fields \cite{Irastorza:2018dyq,Darling:2020plz,Darling:2020uyo} (we work throughout in natural units where $\hbar = c = 1$, such that mass and frequency have the same dimension).

To estimate the resulting radio flux we follow the analysis of Ref.~\cite{Hook:2018iia}, in common with the other observational studies presented in Refs.~\cite{Darling:2020plz,Darling:2020uyo,Foster:2020pgt}. Therein, using a variant of the Goldreich-Julian model \cite{Goldreich:1969sb} assuming the presence of electrons and positrons only, we have a magnetosphere plasma frequency $\omega_{\rm p}\simeq(4\pi\alpha n_{\rm e}/m_{\rm e})^{1/2}$, where $\alpha$ is the fine structure constant and $n_{\rm e}/m_{\rm  e}$ are the electron number density and mass respectively.
Since the axion/photon conversion probability is maximised on resonance, when $m_{\rm a}\simeq \omega_{\rm p}$, we focus on the critical radius where this condition is satisfied,
\begin{eqnarray}
    r_{\rm c}&=& 224 \, \, \text{km}  \big|3 \cos \theta \, {\bf \hat m} \cdot {\bf \hat r} - \cos \theta_{\rm m} \big|^{1/3}  \nonumber\\
 & \times &  \left(\frac{r_{\rm NS}}{10\,{\rm km}} \right)  \left[\frac{B_{0}}{10^{14}\,{\rm G}} \ \frac{1\,{\rm sec}}{P}
\left(\frac{1\,{\rm GHz}}{m_{\rm a}} \right)^2 \right]^{1/3}\,. 
\end{eqnarray}
Here $\theta$ and $\theta_{\rm m}$ are the angles between the NS axis of rotation and respectively, our observational line of sight and the magnetic dipole axis, ${\bf \hat m} \cdot {\bf \hat r} = \cos \theta_{\rm m} \cos \theta + \sin \theta_{\rm m} \sin \theta \cos(\omega t)$. $P$ is the NS period and $\omega=(2\pi/P)$, $r_{\rm NS}$ is the NS radius and $B_0$ is the magnetic field strength at the poles.


Since the NS is relatively close to the Earth ($d\simeq 250$ pc \cite{Posselt:2006ud}), we assume a standard Maxwell-Boltzmann DM velocity distribution with local density $\rho_{\rm DM}^{\infty}=0.45$ GeV\,cm${}^{-3}$ \cite{Bovy:2012tw,Read:2014qva} asymptotically far from the NS surface, which leads to a radio signal with intrinsic linewidth $\Delta f/f\simeq v_0^2$ \cite{Hook:2018iia}, where $v_0\sim 10^{-3}$ is the DM velocity dispersion. 

There are however additional effects which can further broaden our expected signal.
It has, for example, been argued in Refs.~\cite{Battye:2019aco,Battye:2021xvt} that misalignment of the NS rotation axis and the critical surface of axion/photon conversion can result in significant broadening above the intrinsic linewidth, although the conclusions of Ref.~\cite{Foster:2020pgt} differ on this point. More recently this issue has been revisited in more comprehensive detail in \cite{Witte:2021arp}, where it was found that various effects can broaden the signal linewidth by more than an order of magnitude, albeit primarily in the large $|\vec{\omega}|$ regime. However, in the $|\vec{\omega}|\sim 0.1\,{\rm s}^{-1}$ regime we currently operate for NS J0806.4-4123, the additional broadening was found to be minimal, for the fiducial models considered therein.

Therefore we adopt $\Delta f/f\simeq 5\times10^{-6}$, in line with Ref.~\cite{Foster:2020pgt} and the original analysis in Ref.~\cite{Hook:2018iia}.
For the frequency range of our interest the resulting linewidth is then comparable with the 16 kHz MeerKAT channel width, such that the expected signal can be completely contained in three adjacent channels.
The average flux density in channel $i$ is then given by 
\begin{eqnarray}
    \bar{S}_{\nu_i}&=& \frac{F}{\Delta \nu} =
    3.8 \times 10^{-6} \, \, \text{Jy} \, \left(\frac{100\,{\rm pc}}{d} \right)^2\left(\frac{16\,{\rm kHz}}{\Delta \nu}  \right)  \nonumber\\
    & \times &\left(\frac{{{\rm d}\mathcal{P}}/{{\rm d}\Omega}}{5.7\times10^9\rm{W}}\right)
    \int_{\nu_{i,{\rm min}}}^{\nu_{i,{\rm max}}}
    \frac{{\rm d}\nu}{\sqrt{2\pi}\sigma_0}{\rm e}^{-\frac{(\nu-m_{\rm a})^2}{2\sigma_0^2}}\,,
    \label{eq:Sbar}
\end{eqnarray}
where we assume a Gaussian shape of the spectrum with intrinsic width $\sigma_0=5\times10^{-6}\,m_{\rm a}$.
$\nu_{i,{\rm min}}$ and $\nu_{i,{\rm max}}$ are the start and end frequencies of the channel, and $\Delta \nu$ is the channel width. 
The radiated power is given by 
\begin{eqnarray}
\label{eq:dPdOmega}
    \frac{{\rm d}\mathcal{P}}{{\rm d}\Omega} 
    & \simeq & 
    5.7 \times 10^{9} \, \, \text{W} 
    \left( \frac{g_{{\rm a} \gamma \gamma}}{10^{-12}\,\text{GeV}^{-1}} \right)^2 \left(\frac{r_{\rm NS}}{10\,{\rm km}} \right)^{5/2} \left(
    \frac{m_{\rm a}}{\rm GHz} \right)^{4/3} \nonumber\\
    & \times &\left(\frac{B_{0}}{10^{14}\,{\rm G}} \right)^{5/6}\left(\frac{P}{\rm sec} \right)^{7/6}\left( 
    \frac{\rho^{\infty}_{\rm DM}}{0.45\, \text{GeV}\,\text{cm}^{-3}} \right) \left(\frac{M_{\rm NS}}{{\rm M}_{\odot}} \right)^{1/2}\nonumber\\
    & \times &\left( \frac{200\,\text{km}\,\text{s}^{-1}}{v_0} \right)
    \frac{3 \, ({\bf \hat m} \cdot {\bf \hat r})^2 + 1}{\big|3 \cos \theta \, {\bf \hat m} \cdot {\bf \hat r} - \cos \theta_{\rm m} \big|^{7/6}},
\end{eqnarray}
where, following Ref.~\cite{Witte:2021arp} we have re-introduced 
the missing factor of $1/v(r_{\rm c})$ identified in Ref.~\cite{Battye:2021xvt} as being absent from the original expression in Ref.~\cite{Hook:2018iia}. 
Time dependence enters in ``${\bf \hat m} \cdot {\bf \hat r}$'' terms, but lacking sufficient time resolution we average over the NS period to derive our resulting constraint.

It is important to emphasise that this theoretical approach relies on various simplifying assumptions and so has some limitations.
For example, following Refs.~\cite{Leroy:2019ghm,Battye:2021xvt,Witte:2021arp}, incorporating ray tracing of the emitted photons can diminish the estimated total emitted power relative to the analysis of \cite{Hook:2018iia}, corresponding to an $\mathcal{O}(1)$ increase in the minimum detectable $g_{{\rm a}\gamma\gamma}$.
In this sense, our estimated limit can be viewed as less conservative than those arrived at from a full ray-tracing procedure.

A more realistic modelling of the NS magnetosphere and axion/photon conversion in this environment is also needed. 
However, from this perspective our estimate can be regarded as conservative insofar as the analysis of  Ref.~\cite{Millar:2021gzs} suggested that more thorough modelling can provide an enhanced signal strength.
Most recently, it has also been suggested in Ref.~\cite{Bondarenko:2022ngb} that the underlying resonance condition could be extended to arbitrarily light axions, further modifying the underlying theoretical picture.

General relativistic corrections are also absent, although as noted in Ref.~\cite{Leroy:2019ghm} these typically amount to a percent-level correction overall. Reference~\cite{Battye:2021xvt} furthermore suggested that these effects can in fact be helpful insofar as they can counterbalance refractive effects of the magnetosphere, although for our analysis this point is not salient. 

As an overall perspective, it is worth noting that as a relatively novel approach to axion discovery there is at present no analysis in the literature accounting for all these effects.
As emphasised in Ref.~\cite{Witte:2021arp} various aspects of the radio flux estimation, especially the true broadening of the expected signal, are still ongoing topics of debate.
As such, any limits derived therefrom are to be understood within this context.

\begin{figure}[ht]
	\centering
	\includegraphics[width=1\columnwidth]{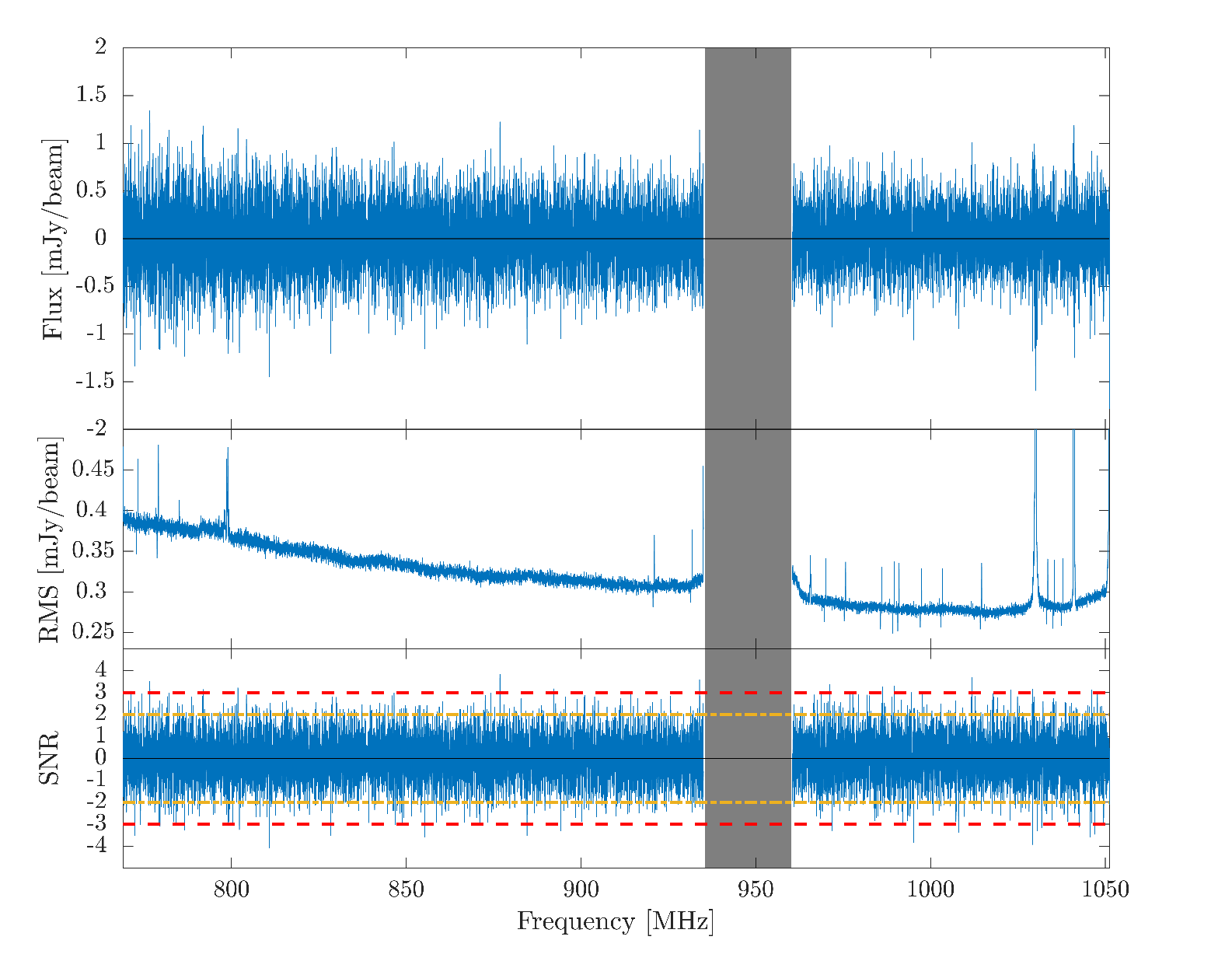}
	\caption{The measured flux density, i.e. spectra of the NS (top), the root-mean-square ({\it rms}; middle) and signal-to-noise ratio (SNR; bottom) as a function of frequency for the 10-hour MeerKAT observation. The grey shaded region is the frequency range $935$-$961\,{\rm MHz}$ most affected by Radio Frequency Interference (RFI). In the bottom panel, we plot the $2\sigma$ C.L. (orange) and $3\sigma$ C.L. (red) limits for visual comparison.}
	\label{fig: flux density}
\end{figure}

\begin{figure}[ht]
	\centering
	\includegraphics[width=1\columnwidth]{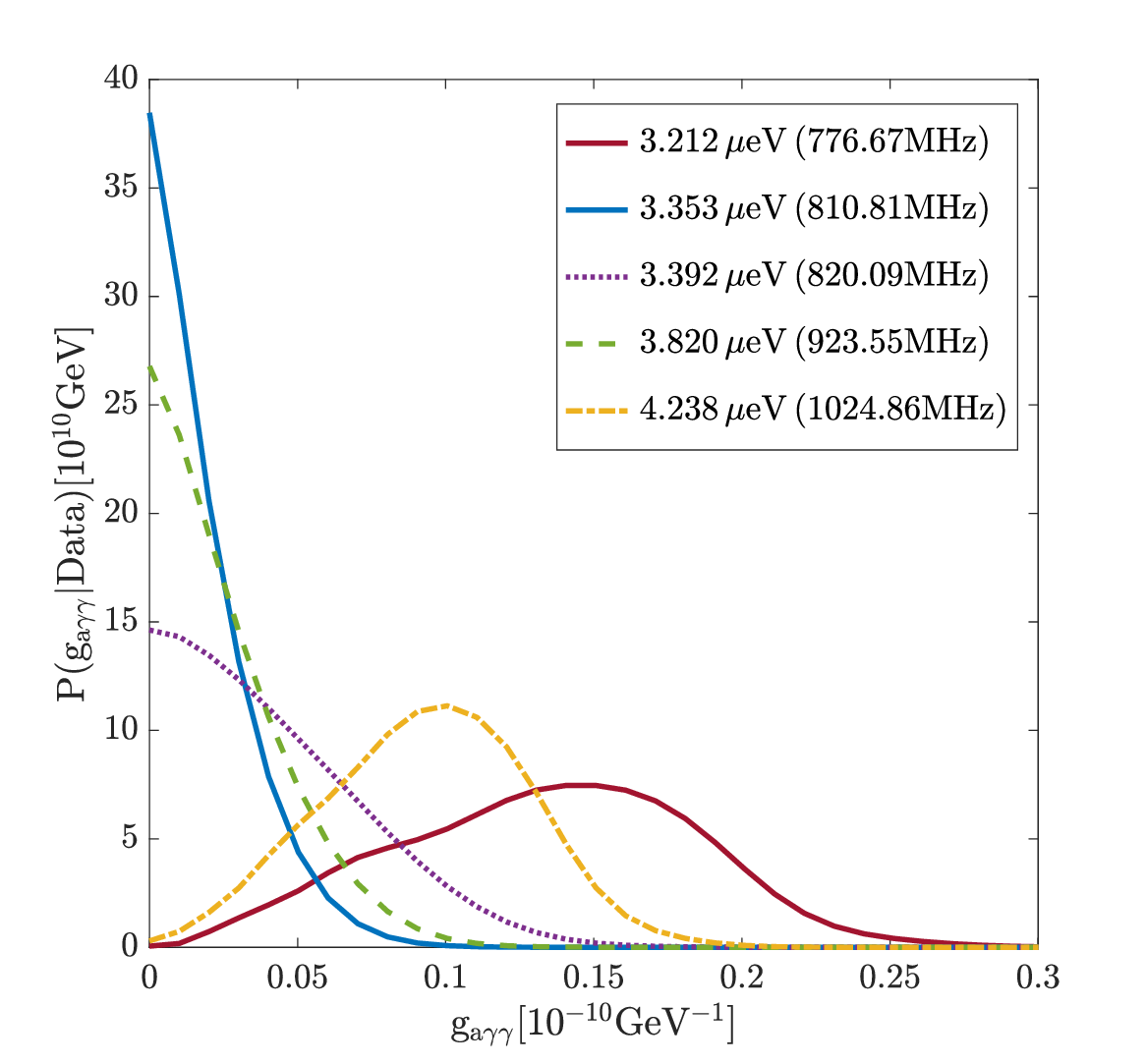}
	\caption{Posterior distribution of the axion coupling constant $g_{{\rm a}\gamma\gamma}$ for $5$ representative axion masses. The $95\%$ C.L. upper limits from these $5$ probability distributions are $(g_{{\rm a}\gamma\gamma}\times 10^{10}~{\rm GeV})<0.213\,(3.212\, \mu\eV)$, $0.054\, (3.353 \,\mu\eV)$, $0.104\,(3.392\, \mu\eV)$, $0.069\,(3.820 \,\mu\eV)$, and $0.149\,(4.238 \,\mu\eV)$. Notice that although being significant apparently, the red line does not give more than $4\sigma$ C.L. detection.}
	\label{fig: likelihood}
\end{figure}

{\bf Data acquisition.} MeerKAT was pointed towards NS J0806.4$-$4123 (${\rm RA}=08^{\rm h}06^{\rm m}23.3471^{\rm s}$, ${\rm DEC}=-41^{\circ}22^{\prime}30.179^{\prime\prime}$, J2000), in six observing epochs of $\sim 100$ minutes, in 2021 (11 March/64 antennas, 14 March/63 antennas, 24 March/61 antennas, 2 April/62 antennas, 4 April/63 antennas, 11 May/60 antennas functional). The complete UHF band ($544$-$1087.983\,{\rm MHz}$ including the tapered edges of the band) was covered in shared-risk commissioning observations. The backend was configured in 32k mode (32768 channels), resulting in a frequency resolution of 16.602 kHz. The integration time was 8s and all (linear) polarisation products were observed and recorded, although the cross-polarisation data were not used for the generation of final data products.
 
To avoid systematic effects associated with the $uv$-coverage we observed the source over three different hour angle ranges. 
In each period, a bandpass calibrator (either J0408$-$6545 or J1939$-$6342, depending on the LST range) was observed for 8 minutes at the beginning and end of the observations, and a gain calibrator (J0828$-$3731) was observed for 2 minutes before each of the 3 target observations, each lasting 26 minutes. 

We made use of the data products provided by the MeerKAT science-data-processor (SDP) pipeline, which can be accessed through the MeerKAT archive interface. 
The SDP pipeline provides raw images (channel maps) from an automated calibration routine, which performs flagging of the data for radio frequency interference (RFI), cross calibration, including bandpass and gain calibrations. 
It then performs a continuum calibration based on the {\sc Obit} data reduction package~\cite{Cotton08}, with two rounds of self-calibration. 

Wide-field effects are dealt with using a faceting approach and wide-band effects are mitigated by using sub-bands. 
The continuum model is then subtracted from the visibilities. 
The continuum-subtracted data are used to produce CLEANed images with Briggs's robust weighting 0 \cite{1995PhDT.......238B} at full frequency resolution ($16.602\,\mathrm{kHz}$) employing a dedicated imaging software created for the usage on GPU units.
We combined the single channel maps from the archive into a data cube and performed a further continuum-subtraction employing a median filter. The original images cover the primary beam size ranging from $104\,\mathrm{arcmin}$ to $208\,\mathrm{arcmin}$ (HPWB), but we made use of an inner fraction of the images only ($14.9$\,arcmin $\times 14.9$\,arcmin) only, sufficient to measure the flux density in the central pixel. The channel maps were convolved with Gaussians such that the resulting synthesized beam size was identical at all frequencies and in all observations, $20.9$\,arcsec $\times$ $13.4$\,arcsec (half-power-beam-width, HPBW) and regridded to the identical pixel size ($3.5\,\mathrm{arcsec}$, fully sampling the synthesized beam).
This gave us 6 data cubes, which were inspected individually to identify potential problems or artifacts in the data, then averaged into a final data cube. 
\\
{\bf Analysis and results.} 
Channels with frequencies below
$769\,{\rm MHz}$ and in the interval of $[935,961]$\,MHz were discarded because the former showed systematic features possibly due to
correlator issues, and the latter due to RFI (Fig.~\ref{fig: flux density}).
Our final 
spectrum covers the ranges of $769$-$935$\,MHz and $961$-$1051$\,MHz, thus occupying $256$\,{\rm MHz} total bandwidth ($16,000$ 
channels, as shown in Fig.~\ref{fig: flux density}). 
Voxel unit is ${\rm mJy/beam}$, a measure of the intensity, also representing the flux density in units 
of ${\rm mJy}$ in the case of point sources.

\begin{figure}[ht]
	\centering
    \includegraphics[width=1\columnwidth]{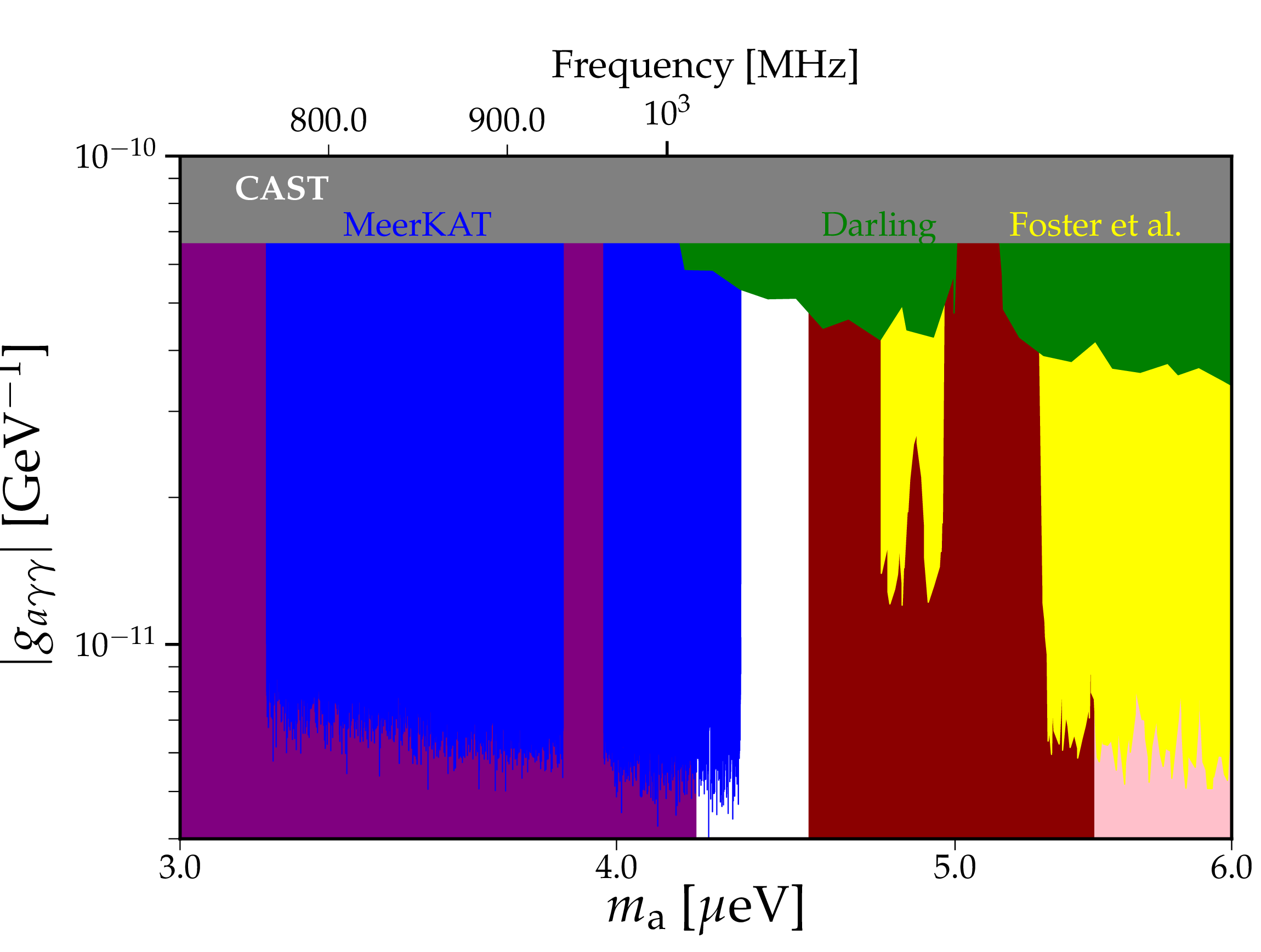}
    \includegraphics[width=1\columnwidth]{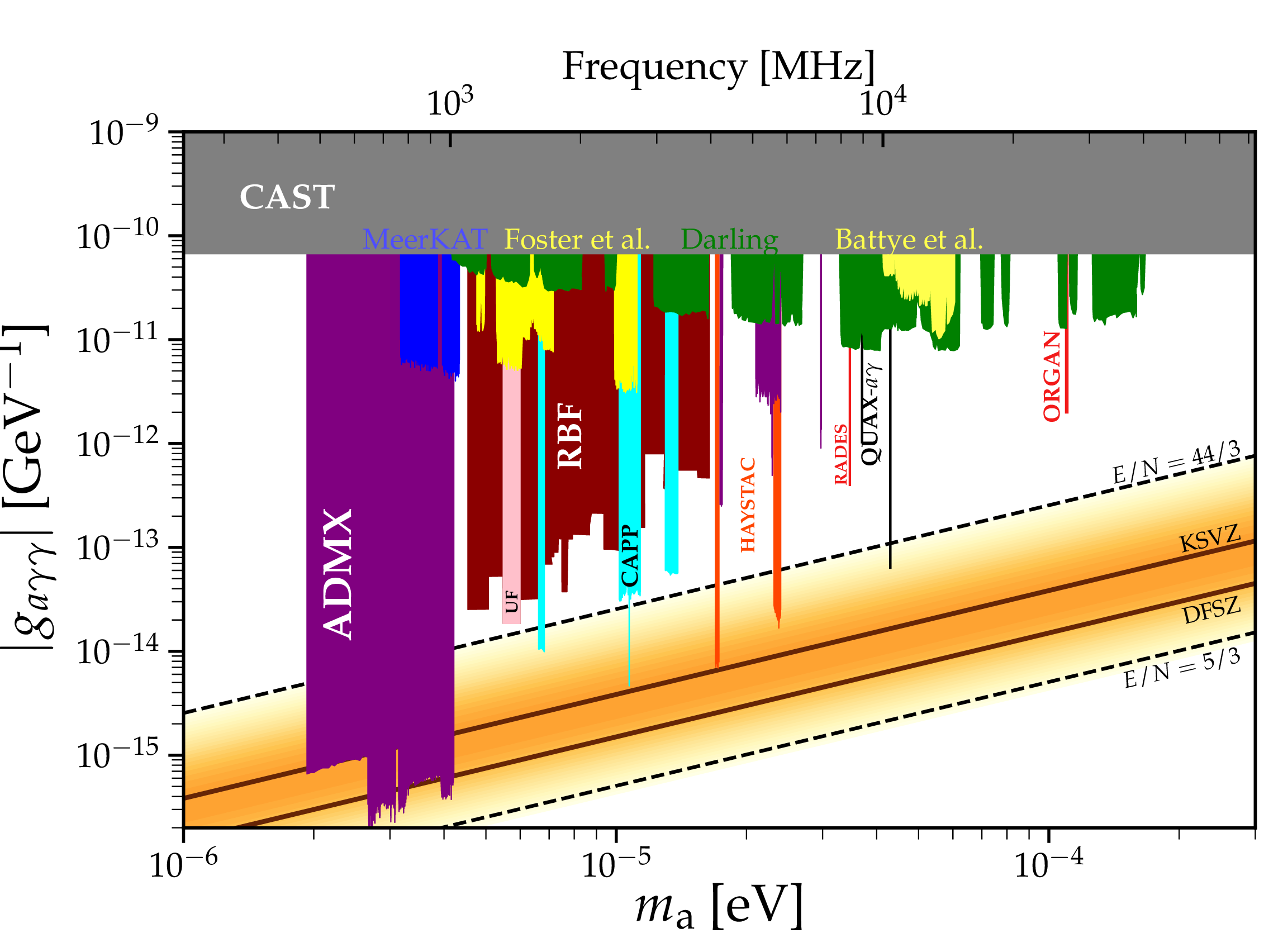}
	\caption{The 95\% C.L. upper limits on $g_{{\rm a}\gamma\gamma}$ derived from $10$-hours of MeerKAT observations of NS J0806.4$-$4123 (blue), compared with other results. The top panel is a cut-out of the MeerKAT-relevant regime from the bottom panel ($3\,{\mu}{\rm eV}<m_{\rm a}<6\,{\mu}{\rm eV}$). Other constraints including those from the same observational technique for various targets (green, yellow) \cite{Foster:2020pgt,Darling:2020plz,Darling:2020uyo,Battye:2021yue}, as well as laboratory experiments \cite{CAST:2017uph,ADMX:2019uok,Bartram:2021ysp,ADMX:2021abc,Hagmann,DePanfilis,CAPP:2020utb,HAYSTAC:2020kwv,CAST:2021add,Alesini:2020vny,McAllister:2017lkb}.
	The orange shaded region in the lower panel represents the parameter space of typical QCD axion models, such as the well-known DFSZ/KSVZ models, where the mass and couplings are inversely correlated.
	Figure was produced using the {\tt AxionLimits} code~\cite{O'Hare:2020}.}

	\label{fig: axion constraint}
\end{figure}

We now analyze the cleaned cube to extract the flux density and search for potential axion signal. 
The central pixel coincides with the 
pulsar's location and its value reflects the source's flux density ($d_{\nu_{i}}$). 
We adopt an ``Aperture Photometry'' method to characterize the background fluctuations
\cite{Planck-unbound,Li2018}. We draw two circles that cover the center pixel with radii 
$2.25$\,arcmin and $6$\,arcmin, respectively.
The outer circle is sufficiently large so that 
the ring area can represent the background fluctuations. 
We then compute the mean ($\mu_{\nu_i}$) 
and root-mean-square ({\it rms}; $\sigma_{\nu_{i}}$) values of all pixels contained in the region 
enclosed by the two circles (roughly $28,500$ pixels). 
We treat ($\mu_{\nu_{i}}$, $\sigma_{\nu_{i}}$) 
as the mean and {\it rms} of the background continuum, so any possible detection of axion signal
in the pulsar location should have an excess compared with the background value (see also Eq.~(\ref{eq:log-likelihood})). 
We compare the measured and the expected {\it rms} for five channels using the SEFD provided in the 
MeerKAT description in the radiometer 
equation~\cite{1999ASPC..180..171W}, to then multiply with a typical correction factor of 1.45. This ``fudge factor'' is typical for MeerKAT, accounting
for Robust-0 weighting (as compared to Natural weighting, for which the radiometer equation is valid). The measured {\it rms} matches the 
theoretical prediction well, confirming the integrity of the data.

The mean and {\it rms} values for each channel are plotted in the top and middle panels of Fig.~\ref{fig: flux density}. One can see that the measured flux density varies for each frequency channel but the mean is about zero across all frequency bands. 
In addition, the absolute flux value decreases slightly from low to high frequency bands. This trend is exactly captured in the middle panel, in which we plot the {\it rms} value ($\sigma_{\nu_i}$) for all channels. 
The {\it rms} decreases from $0.4$ to $0.28\,{\rm mJy/beam}$ and therefore strengthens the constraint at higher frequencies. 
The trend can be understood as a consequence of a lower system temperature in the UHF band towards higher frequencies, although the amplitude of the decrease is slightly higher than expected. 

In the bottom panel of Fig.~\ref{fig: flux density} we plot the signal-to-noise ratio, defined as the ratio between flux density and the corresponding channel's {\it rms}, and overlaid the $2\sigma$ (orange) and $3\sigma$ (red) confidence level (C.L.) limits for visual comparison. It is clear that the SNR of each channel varies, but is centered at zero. In addition, we checked that there are $40$ channels with SNR exceeding $\pm3\sigma$, and only one channel exceeding $\pm 4\sigma$ SNR. Since there are $16,000$ channels in total, the expected numbers of channels that can exceed $3\sigma$ and $4\sigma$ C.L. due purely to the noise fluctuation are $16,000 \times 0.27\%=43.2$, and $16,000\times 0.0063\%=1.01$. These numbers are in perfect consistency with the measurements. No peak exceeding $5\sigma$ has been found in our data. The statistics indicates that the rare peaks we have found are within the allowance of noise fluctuations. We therefore regard the measured flux density as consistent with noise in the radio continuum background, and derive upper limits on the Primakoff coupling constant between axions and photons from the data. The worry might arise whether the signal might have been reduced by the clipping algorithm of the data reduction pipeline. However, since the source is at the phase centre, its potential signal in one channel would be constant over all times and hence either be clipped in the majority of all visibilities (reducing in a missing channel) or just in the minority of all channels (resulting in the preservation of the signal). In the frequency ranges 769-935 MHz and 961-1051 MHz as analysed in this work no signal was totally clipped, which suggests that the non-detection of the signal is not due to data processing. A possible test by ingesting an artificial source before data processing was not done because the pipeline was not directly available to the authors.

A Bayesian approach is employed to obtain the constraints on axion parameters. The likelihood function for axions with mass $m_{\rm a}$ and coupling $g_{{\rm a}\gamma\gamma}$ is given by
\begin{equation}
\mathcal{L}\left(m_{\rm a}, g_{{\rm a}\gamma\gamma}\right)=\prod_{i=1}^{N_{\rm ch}}
\frac{1}{\sqrt{2\pi}\sigma_{\nu_{i}}}\exp\left[-\frac{\left(d_{\nu_i}-\mu_{\nu_i}-\bar{S}_{\nu_i}
(m_{\rm a}, g_{{\rm a}\gamma\gamma})\right)^{2}}{2\sigma^{2}_{\nu_{i}}} \right],
\label{eq:log-likelihood}
\end{equation}
where the product runs over all channels. We substitute the theoretical $\bar{S}_{\nu_i}$ value calculated via Eq.~(\ref{eq:Sbar}) into Eq.~(\ref{eq:log-likelihood}) by inserting $r_{\rm NS}=10\,{\rm km}$, $M_{\rm NS}={\rm M}_\odot$, $P=11.4\,{\rm s}$, $d=250\,{\rm pc}$, $B_0=2.5\times10^{13}\,{\rm G}$~\cite{Posselt:2006ud,2009ApJ...705..798K} and adopting $\rho_{\rm DM}^{\infty}=0.45\,{\rm GeV}\,{\rm cm}^{-3}$ \cite{Bovy:2012tw,Read:2014qva} and $v_0=200\,{\rm km\,{\rm s}^{-1}}$. 
The velocity dispersion is close to the canonical value usually adopted in direct detection 
experiments (e.g., \cite{XENON:2017vdw}). For other choices of parameters, the results can be easily 
scaled according to Eq.~(\ref{eq:dPdOmega}).
We then sample $g_{{\rm a}\gamma\gamma}$ in the range $[0,2\times 10^{-10}]\,{\rm GeV}^{-1}$ and marginalize over $\theta,\,\theta_{\rm m}$ in the ranges $[0,\pi]$, $[0,\pi/2]$ respectively to obtain the posterior distribution of $g_{{\rm a}\gamma\gamma}$ with a flat prior in its sampling range.

Figure~\ref{fig: likelihood} shows the posterior probability distributions of $g_{{\rm a}\gamma\gamma}$ for five selected axion masses. 
One can see that because of differing SNRs, the distribution functions are centered at different values of $g_{{\rm a}\gamma\gamma}$. 
However, even the distribution function with the highest significance ($m_{\rm a}=3.212\,{\rm \mu eV}$) does not favour non-zero $g_{{\rm a}\gamma\gamma}$ at more than $4\sigma$. 
We therefore regard all of the distributions as providing upper limits of $g_{{\rm a}\gamma\gamma}$. Our final $95\%$ C.L. limits of $g_{{\rm a}\gamma\gamma}$ along with other observational and experimental constraints are shown in Fig.~\ref{fig: axion constraint}.

{\bf Conclusion.}
From 10 hours of MeerKAT observations we have explored the conversion of DM axions into radio-frequency photons in the magnetosphere of the isolated neutron star J0806.4$-$4123, constraining the axion/photon coupling to be $g_{{\rm a}\gamma\gamma}\lesssim 9.3 \times 10^{-12}\,{\rm GeV}^{-1}$ at $95\%$ C.L. in the mass range of $3.18$-$4.35\,\mu$eV. This result provides the strongest known constraints in the mass range of $4.20$-$4.35\,\mu$eV, which is not covered by ADMX~\cite{ADMX:2021abc}. Compared with the other constraints derived from the NS magnetospheric axion/photon conversion, we have extended the studied window of axion to lower masses thanks to the UHF receiver of MeerKAT, and have reached better performance at overlapping frequencies due to the high sensitivity of MeerKAT in this range.

These constraints can be improved via future observations in several ways. 
Given the time-dependent nature of the anticipated signal induced by NS rotation, an increased sensitivity can in principle be achieved by exploiting time-series data, as discussed in \cite{Hook:2018iia}.
More observations of other highly magnetised sources can also lead to tighter constraints. 
In the future, the analysis procedure, especially the possible confusion of signal with RFI may also
be investigated through injecting synthetic signals either in the experiment or the simulation. On the theoretical side there are also a number of aspects, recently explored in literature \cite{Witte:2021arp,Battye:2019aco,Leroy:2019ghm,Battye:2021xvt,Millar:2021gzs}, which can be 
incorporated to improve the accuracy of our constraints. 
These include corrections due to gravitational and relativistic effects, ray tracing of the emitted photons and a more sophisticated modelling of magnetospheric plasma effects.


\begin{acknowledgments}
The MeerKAT telescope is operated by the South African Radio Astronomy Observatory, which is a facility of the National Research Foundation, an agency of the Department of Science and Innovation. 
Apart from the raw data provided by SARAO, this study relies on the processed data products as provided by the Science Data Processor pipeline of MeerKAT. 
We also acknowledge use of the \texttt{NumPy} and \texttt{SciPy} scientific computing packages \cite{harris2020array, 2020SciPy-NMeth}.
NH is supported by the National Natural Science Foundation of China (NSFC) under Grant No. 12150410317.
HC is supported by the South African Department of Science and Innovation, the National Research Foundation through a SARChI's South African SKA Fellowship within the SARAO Research Chair held by RC Kraan-Korteweg, and the Key Research Project of Zhejiang Lab (No. 2021PE0AC03).
YZM is supported by the National Research Foundation of South Africa under Grant No. 120385 and No. 120378, NITheCS program ``New Insights into Astrophysics and Cosmology with Theoretical Models confronting Observational Data'', and National Natural Science Foundation of China with Project No. 12047503. QY is supported by the Key Research Program of Chinese Academy of Sciences (No. XDPB15) and the 
Program for Innovative Talents and Entrepreneur in Jiangsu. 
FD is supported by the National Natural Science Foundation of China (NSFC) under Grant No. 11873094. 
YC acknowledges the support from the NSFC under Grant No. 12050410259, and Center for Astronomical Mega-Science, Chinese Academy of Sciences, for the FAST distinguished young researcher fellowship (19-FAST-02), and MOST for the Grant no. QNJ2021061003L. 
RD is supported in part by the National Key R\&D Programme of China (2021YFC2203100).
We acknowledge the use of China SKA Regional Center prototype system at Shanghai Astronomical Observatory, funded by the National Key R\&D Programme of China (No. 2018YFA0404603) and Chinese Academy of Sciences (No. 114231KYSB20170003).
\end{acknowledgments}

\bibliography{Ref}

\vspace{-7mm}

\end{document}